# Multidimensional Data Structures and Techniques for Efficient Decision Making

MADALINA ECATERINA ANDREICA
The Bucharest Academy of Economic Studies
6, Romana Square, District 1, Bucharest
ROMANIA
madalina.andreica@gmail.com

MUGUREL IONUT ANDREICA
Computer Science and Engineering Department
Politehnica University of Bucharest
Splaiul Independentei 313, sector 6, Bucharest
ROMANIA
mugurel.andreica@cs.pub.ro   https://mail.cs.pub.ro/~mugurel.andreica

NICOLAE CATANICIU
National Scientific Research Institute for Labor and Social Protection
6-8 Povernei Str., District 1, Bucharest
ROMANIA
ncataniciu@incsmps.ro

*Abstract:* - In this paper we present several novel efficient techniques and multidimensional data structures which can improve the decision making process in many domains. We consider online range aggregation, range selection and range weighted median queries; for most of them, the presented data structures and techniques can provide answers in polylogarithmic time. The presented results have applications in many business and economic scenarios, some of which are described in detail in the paper.

*Key-Words:* - Decision making, Multidimensional data structures, Risk management, Range aggregation query, Range selection query, Range minimum query, Weighted median.

## 1 Introduction

The process of decision making is both a permanent necessity and a challenge in many economic and industrial fields, like risk management, banking, selection of financing means, such as leasing or credit, bonity analysis of clients, e-commerce, operational research, and many others. The importance of the decision making process has been confirmed by the large number of publications which develop and propose efficient decision making techniques. These techniques can be classified into several broad categories, such as those handling certain and complete information, those using uncertain data and objectives, and those based on risk assessment. Some of the best known decision making optimization methods are multi-attribute and multi-objective decision making [1], fuzzy decision rules [2] and dynamic programming [3].

In order for the decision making process to obtain significant results, we need two factors: a good optimization technique and accurate input values. Although a lot of effort has been directed towards developing highly efficient decision making optimization methods and models, the process of obtaining accurate input information as quickly as possible seems to have been mostly overlooked. This paper presents several techniques and data structures for obtaining aggregate information efficiently from a large database, which can later be used as input data in a decision process.

The main scenario in which we consider the implementation of the presented techniques is the following: A large database, modeled as a multidimensional data cube, is available on a central server, which can be accessed by multiple clients (e.g. economic agents). Every client can ask at any time for the computation of some aggregate information (e.g. min, max, sum, average, median) over several (large) portions of the database. Different clients may be interested in different parts of the database. Using standard techniques in such cases would lead to unacceptably high response times, especially in the case of many clients accessing the database simultaneously. On the other hand, the techniques we propose in this paper lead to response times which are several orders of magnitude better than the common methods. The considered scenario occurs, for instance, in the case of large companies or banks which have all of their data stored in a data center which is accessed by their geographically spread headquarters and subsidiaries.

The rest of this paper is structured as follows. Section 2 presents several use cases for the implementation of the techniques and data structures which are introduced in Section 3. In Section 4 we discuss related work and we conclude.

## 2 Use Cases

In this section we present several situations where our techniques can be used successfully. Let's consider the case of a large retail company which handles thousands of transactions per day (both in department stores and over the Internet). The information associated to each transaction (e.g. date and time, purchase price, quantity, personal information about the customer, customer's answers to relevant surveys) is carefully stored in a central database. Periodically, the company's managers have to decide upon the price, quantity and (types of) products which should keep being sold or which should be taken off the market or improved, or if any new products should be launched. In this decision making process, they need quick access to aggregate information like: the total amount of sales of certain types of products to customers of a certain age and income range; the degree of satisfaction of each consumer target group which acquired products with certain characteristics; the percentage of the sales of each product from the total amount of sales; the most likely characteristics (age and income) of the highest paying consumers from a given age and income range.

Let's consider now a financial consulting company (e.g. specialized on risk assessment and management) which owns a private database with information regarding the outcomes of several strategic decisions made by multiple companies over a large period of time. A company is characterized by multiple parameters, such as total income, profit, investments, financial sources, market share, supply characteristics on the market, and so on. A strategic decision is also characterized by several specific parameters. When advising a client, the consulting company may compute an aggregate of the outcomes of the strategic decisions made by the competing companies with similar characteristics to the customer's company (i.e. using intervals containing the parameter values of the customer's company).

Another example consists of a bank which needs to perform a bonity analysis of its clients soliciting different credits. This is a crucial decision making process, especially in the context of the present global financial crisis. This is why the banks should use as much available data as possible and obtain the results in the most efficient manner.

On the other hand, there are several situations when the access to available data should be significantly improved. For instance, let's consider the case of a leasing company that wants to access some aggregate information regarding the financial status of several competing leasing companies on the market during a specified period of time. At the moment, the company can only obtain the financial balance sheets from the Ministry of Finance, from which it can compute the aggregate information by itself. It would be more efficient if the Ministry provided an e-government service for retrieving the required aggregate information automatically. All the presented use cases can benefit from using the techniques and data structures presented in this paper, as most of the desired information can be easily expressed as range aggregate queries.

## 3 Multidimensional Data Structures

### 3.1 Multidimensional Range Minimum Queries

The RMQ technique [4] is very versatile and can be extended to multi-dimensional arrays. Each query asks for the minimum (maximum) value in a range $[r(1,1),r(1,2)] \times \ldots \times [r(d,1),r(d,2)]$ of a d-dimensional array $a$. W.l.o.g., we assume that the array contains $n$ cells in every dimension (and, thus, $n^d$ cells overall). We will consider the most general case, in which the dimensions can be split into $e \leq d$ groups (*group(j)*=the group of dimension *j*). Each group $g$ ($1 \leq g \leq e$) contains a base dimension $bd(g)$. The length of the interval of any query for every dimension $j$ is always equal to ($f(j)$ x the length of the query interval in dimension $bd(group(j))$), with $f(j) \geq 1$ ($f(j)=1$ for $j=bd(group(j))$). Since all the dimensions except the base dimensions are "linked" to the base dimension in their group, we will compute the following values: $m(c(1), \ldots, c(d), k(1), \ldots, k(e))$ = the minimum value in a range where $r(j,1)=c(j)$ and the length of the query interval in dimension $bd(g)$ is $2^{k(g)}$. We will consider the sequences $(k(1), \ldots, k(e))$ in increasing lexicographic order. When at least one $k(*)$ value is larger than $0$, we can use eq. (1), where $q(u)=1$, if $(k(group(u))>0)$, and $0$, otherwise.

$$m(c(1),...,c(d),k(1),...,k(e)) = \min_{\substack{(s(1),...,s(d))\in\{0,1\}^d \\ s(u)\leq q(u), 1\leq u\leq d}} \{$$
$$m(c(u) + q(u) \cdot s(u) \cdot f(u) \cdot 2^{k(group(u))-1}, ...(1 \leq u \leq d)..., \quad (1)$$
$$q(bd(1)) \cdot (k(1)-1),...,q(bd(e)) \cdot (k(e)-1)\}$$

A better way to compute $m(c(1), \ldots, c(d), k(1), \ldots, k(e))$ is to select just one value $k(j)>0$, set $q(j)=1$, replace $q(bd(u)) \cdot (k(u)-1)$ by $k(u)$ in eq. (1) (for $u \neq j$), and instead of considering all the tuples $(s(1), …, s(d))$ in eq. (1), we can consider only the tuples $(s(1), …, s(d))$ with $s(p)=0$ or $1$, only if $group(p)=j$ (and $0$, if $group(p) \neq j$). When all the $k(*)$ values are $0$, we distinguish between two cases. If every group contains only one dimension or all the $f(j)$ values of the non-base dimensions $j$ are $1$, then $m(c(1), …, c(d))=a(c(1), …, c(d))$. Otherwise, the problem is reduced to computing the minimum value in a range $[r(1,1),r(1,2)] \times … \times [r(d,1),r(d,2)]$, where $r(j,1)=r(j,2)$ if $j$ is a base dimension in its group, or $r(j,2)=r(j,1)+f(j)-1$, if $j$ is not a base dimension. We can handle this case as $O(n^d)$ (d-e)-dimensional RMQ queries (one for every tuple of coordinates). The (d-e) dimensions are the remaining (d-e) non-base dimensions, grouped as follows. We remove the base dimension $bd(g)$ from each group $g$ and set as the new base dimension of $g$ the non-base dimension $j'$ in $g$ with the minimum value $f(j')$ (if it exists). The new $f(*)$ values of the non-base dimensions $j''$ from a group $g$ will be $f(j'')/f(j')$ (if this is not an integer, we place $j''$ in a separate group). In order to compute the minimum value in a given range $[r(1,1),r(1,2)] \times … \times [r(d,1),r(d,2)]$, we compute the values $k'(j)=floor(log_2(r(j,2)-r(j,1)+1))$ and set $k(g)=k'(bd(g))$ for every group $g$. The answer is:

$$\min_{(s(1),…,s(d)) \in \{0,1\}^d} \{m(r(u,1) + s(u) \cdot (r(u,2) - f(u) \cdot 2^{k(group(u))}$$
$$+1-r(u,1)),…(1 \leq u \leq d)…,k(1),…,k(e)\} \quad (2)$$

The preprocessing time complexity is at most $O(2^d \cdot n^d \cdot log^d(n))$. A query is answered in $O(2^d)$ time. It is obvious that the *min* function can immediately be replaced by *max*, obtaining identical results for range maximum queries.

## 3.2 Multidimensional Range Aggregate Queries

In this section we consider several alternatives for multidimensional range aggregate queries over a (dynamic) data cube, in which the aggregate function is invertible. The data cube has a fixed number $d$ of dimensions and has $m(j)$ entries in every dimension $j$ ($1 \leq j \leq d$); the entries are numbered from $1$ to $m(j)$. A cell of the data cube has coordinates $(c(1), …, c(d))$ ($1 \leq c(j) \leq m(j)$; $1 \leq j \leq d$) and has a value $Cube(c(1), …, c(d))$. A query $Q(a(1), b(1), …, a(d), b(d))$ consists of computing an aggregate (e.g. *sum*, *product*, *xor*) over the values of the cells $(c(1), …, c(d))$ whose coordinates are in the range: $a(j) \leq c(j) \leq b(j)$ ($1 \leq j \leq d$). An update $U(u, c(1), …, c(d))$ modifies the value of the cell $(c(1), …, c(d))$ by the value $u$ (e.g. it increases/multiplies it by or sets it to $u$). We are interested in supporting both types of operations efficiently. The straight-forward solution is to update every cell in $O(1)$ time (just change the value of $Cube(c(1), …, c(d))$) and compute the aggregate in $O(n^d)$ time ($n=max\{m(1), …, m(d)\}$), by traversing every cell in the given range. For the approaches we present next, we will assume that the query has the form $Q(b(1), …, b(d))$ and asks for an aggregate of the values of all the cells in the range $[1,b(1)] \cdot … \cdot [1,b(d)]$ (i.e. a *prefix subcube* of the data cube). The aggregate of any arbitrary range can be computed as a "sum" of the aggregates of $O(2^d)=O(1)$ prefix subcubes [6]. Thus, considering only prefix subcube queries is enough. In fact, we can compute a prefix "sum" cube, with which we can compute the "sum" (aggregate) of a prefix subcube in $O(1)$ time. However, when an update occurs, we need to recompute the prefix "sum" cube (which takes $O(n^d)$ time). Another solution is based on constructing a multi-dimensional binary indexed tree [7], which supports updates and prefix subcube queries in $O(log^d(n))$ time each (with only $O(n^d)$ memory consumption). Yet another solution is based on using a multidimensional block partitioning [5], which supports updates in $O(n^{d/2})$ time and queries in $O(1)$ time. In this section we propose a novel data structure, which supports both updates and queries in $O(n^{d/4})$ time. The main idea is to divide the dimensions into two disjoint sets: the first set consists of $q$ dimensions and the second set consists of the remaining $d-q$ dimensions. The $O(n)$ entries in every dimension $j$ are split into $O(n/k)$ blocks (where $k$ is a function of $n$) of (approximately) $k$ consecutive entries each. The blocks are numbered starting with $1$, as they appear in increasing order (in increasing order of the entries they contain). Moreover, for every entry $p$ in the $j^{th}$ dimension we know the block number of the block which contains the entry, $blk(j,p)$. For every combination $(x(1), x(2), …, x(q))$, where $x(j)$ is either an entry in dimension $j$ or a block in dimension $j$ (thus, there are $O((n+n/k)^q)$ tuples overall), we maintain a multidimensional block partition $BP(x(1), …, x(d))$ of the remaining $d-q$ dimensions.

When an update $U(u, c(1), …, c(d))$ occurs, we proceed as follows. We consider all the tuples $(x(1), …, x(q))$, where $x(j)$ is either an entry $x(j)=c(j)$, or a block $x(j)=blk(j, c(j))$ ($1 \leq j \leq q$; we consider $O(2^q)=O(1)$ tuples overall). For each such tuple, we will update its multidimensional block partition $BP(x(1), …, x(d))$. We consider all the tuples $(y(q+1), …, y(d))$, where $y(j)$ is either an entry $y(j) \geq c(j)$ with $blk(j, y(j))=blk(j, c(j))$, or a block $y(j)>blk(j, c(j))$ ($q+1 \leq j \leq d$; we consider $O((k+n/k)^{d-q})$ such tuples). We update the cell $(y(q+1), …, y(d))$ of $BP(x(1), …, x(q))$ by $u$ (e.g. add $u$ to it, multiply it by $u$, xor it with $u$). Thus, an update takes $O((k+n/k)^{d-q})$ time.

In order to find the answer to a query $Q(b(1), …, b(d))$, we consider all the tuples $(x(1), …, x(q))$, where $x(j)$ is either an entry $x(j) \leq b(j)$ with $blk(j, x(j))=blk(j, b(j))$, or a block $x(j)<blk(j, b(j))$ (we consider $O((k+n/k)^q)$ such tuples). For each such tuple $(x(1), …, x(q))$, we will compute $Qagg(x(1), …, x(q), b(q+1), …, b(d))$=the aggregate of the values of the following entries $(y(q+1), …, y(d))$ of $BP(x(1), …, x(q))$: $y(j)$ is either an entry $y(j)=b(j)$ or a block $y(j)=blk(j, b(j))$ (thus, we consider $O(2^{d-q})=O(1)$ tuples $(y(q+1), …, y(d))$). The answer to the query is the aggregate of the values $Qagg(x(1), …, x(q), b(q+1), …, b(d))$, over all the considered tuples $(x(1), …, x(q))$. Thus, a query takes $O((k+n/k)^q)$ time. We can choose the parameters $k$ and $q$ as we see fit, possibly according to the expected ratio between queries and updates. Note that when $k=n^{1/2}$ and $q=d$, we obtain $O(1)$ update time and $O(n^{d/2})$ query time; if $q=0$ we move to the other end of the spectrum, where an update takes $O(n^{d/2})$ time and a query takes $O(1)$ time. If both queries and updates are equally probable, we can choose $q=d/2$ (and $k=n^{1/2}$), which, for even $d$, leads to $O(n^{d/4})$ query and update times.

### 3.3 Multidimensional Medians

We consider $n$ points on the real line, each having an x-coordinate $x(i)$ and a weight $w(i)$ ($1 \leq i \leq n$). We want to place (at most) $K$ intervals of fixed length $L$, such that the total sum of distances from the given points to the intervals is minimum. The distance from a point $x(i)$ to an interval $[a,b]$ is defined as: $0$, if $(a \leq x(i) \leq b)$; $min\{|x(i)-a|, |x(i)-b|\}$, otherwise. The problem of determining point K-medians (with fixed length $L=0$) has been studied before [8, 9] and optimal $O(n \cdot K)$ algorithms have been given. We will extend those algorithms to the $L>0$ case. In order to facilitate the solution to this problem, we will add a new point at coordinate $x(i)+L$ (with weight $0$), for every initial point $x(i)$ ($1 \leq i \leq n$), thus obtaining $n'=2 \cdot n$ points. We consider the points sorted in ascending order of their coordinates: $x(1) \leq x(2) \leq … \leq x(n')$. For each point $i$ we compute the value $pleft(i)$=the smallest index such that $x(i)-x(pleft(i)) \leq L$. These values can be computed in $O(n')$ time (if the points are sorted), with a sliding window-type algorithm. In an optimal solution, the medians have the right endpoint positioned at the x-coordinate of some point. We will compute the following values: $D_{min}(i,j,0)$=the minimum total sum of distances of the first $i$ points after placing $j$ intervals and the rightmost interval has its right endpoint at $x(i)$; $D_{min}(i,j,1)$=the minimum total sum of distances of the first $i$ points after placing $j$ intervals and the rightmost interval has its left endpoint at an x-coordinate less than or equal to $x(i)$. We have the following equations: $D_{min}(0,j,0)= D_{min}(0,j,1)=0$ and $D_{min}(i,0,0)=D_{min}(i,0,1)=+\infty$, for $i>0$. For $i>0$ and $j>0$, we have:

$$D_{\min}(i,j,0) = \min_{0 \leq p < pleft(i)} \left\{ D_{\min}(p, j-1,1) + \sum_{q=p+1}^{pleft(i)-1} w(q) \cdot (x(i) - L - x(q)) \right\} \quad (3)$$

$$D_{\min}(i,j,1) = \min_{0 \leq p \leq i} \{D_{\min}(p,j,0) + \sum_{q=p+1}^{i} w(q) \cdot (x(q) - x(p)) \quad (4)$$

These equations are very similar to the equations used in [9] and, thus, the $O(n \cdot K)$ solution presented there can be easily adapted. The main difference consists of the fact that, when computing $D_{min}(i,j,0)$, we only have in the appropriate sorted double-ended queue (deque) values corresponding to candidate positions $p$ ($0 \leq p < pleft(i)$), instead of the whole range $[0, i-1]$; the value corresponding to a position $i'$ is inserted into the deque only when we reach a position $i$, such that $i' < pleft(i)$.

A well-known case of the interval K-median problem is where $K=1$. Like in the general case, we add the $n$ extra points $x(i)+L$ with zero weights, obtaining $n'=2 \cdot n$ points. We will now slide the interval from left to right, placing its right endpoint at every point. While sliding the interval, we will maintain four values: $wd_{left}$, $wd_{right}$ and $w_{left}$, $w_{right}$, representing the total weighted distance of the points on the left (right) side of the interval and the total weight of these points. We assign a type $type(i)$ to every point $i$: $type(i)=0$ if it is one of the original points and $type(i)=-j$ if point $i$ was added as the extra point corresponding to point $j$ (thus, $x(i)=x(j)+L$). Afterwards, we sort the points in increasing order of their coordinates and renumber them in this order (including the negative $type(i)$ values). The pseudocode below presents the algorithm. We consider the points sorted in increasing order of their coordinates:

**Interval-1-Median():**
*right=1; $wd_{left}=w_{left}=0$; $w_{right}$=sum of the values $w(i)$ ($2 \leq i \leq n$)*
*wdm=wdc=$wd_{right}$=sum of the values $(x(i)-x(1)) \cdot w(i)$ ($2 \leq i \leq n$)*
**for** *i=2* **to** *n'* **do** *{*
  *$wd_{right}=wd_{right}-w_{right} \cdot (x(i)-x(i-1))$; $w_{right}=w_{right}-w(i)$*
  *$wd_{left}=wd_{left}+w_{left} \cdot (x(i)-x(i-1))$*
  **if** *(type(i)<0)* **then** *{ $w_{left}=w_{left}+w(-type(i))$ }*
  *wdm=min{wdm, $wd_{right}+wd_{left}$}*
  *wdc=min{wdc, max{$wd_{right}$, $wd_{left}$}} // only for L=0 }*

*wdm* is the minimum sum of weighted distances, corresponding to the interval 1-median. In the multidimensional case we are given $n$ d-dimensional weighted points, at coordinates $(x(i,1),…,x(i,d))$ and with weights $w(i)$ ($1 \leq i \leq n$). We want to place $1$ hyper-rectangle with side lengths $L(j)$, $1 \leq j \leq d$ ($L(j)$ is the side length in dimension $j$) such that the sum of the distances ($L_1$) from the points to the hyper-rectangle is minimized. The distance from a point $i$ to a hyper-rectangle whose lower corner is at $(xr(1), …, xr(d))$ is:

$$\sum_{j=1}^{d} \begin{cases} 0, \text{if } xr(j) \leq x(i,j) \leq xr(j) + L(j) \\ w(j) \cdot (xr(j) - x(i,j)), \text{if } x(i,j) < xr(j) \\ w(j) \cdot (x(i,j) - (xr(j) + L(j))), \text{if } x(i,j) > xr(j) + L(j) \end{cases} \quad (5)$$

We can decompose the problem into $d$ one-dimensional problems. For each dimension $j$, we can find the coordinate $xr(j)$ independently of the other dimensions, by solving a 1D interval 1-median problem, considering points $i$ ($1 \leq i \leq n$) with x-coordinates equal to $x(i,j)$ and having weights $w(i)$, and the length of the interval median is $L(j)$.

A situation which arises often in multidimensional data analysis is that of computing the median of a subrange of the given multidimensional data (using the $L_1$ metric). Let's consider the 1D case first and see how we can support efficient range weighted median queries, i.e. $Q(i,j)$=find the sum of weighted distances from the median of the points $x(r)$, with $i \leq r \leq j$, to all the points in the range. We will show that, with appropriate preprocessing ($O(n)$), we will be able to answer every such query in $O(log(n))$ time. We define $wsum(i,p)$ ($i \leq p$), the sum of the weights of all the points between $i$ and $p$ (inclusive), $wdsumLR(i,p)$=the sum of the weighted distances from every point $j$ ($i \leq j \leq p$) to point $p$, and $wdsumRL(i,p)$=the sum of the weighted distances from every point $j$ ($i \leq j \leq p$) to point $i$. With $O(n)$ preprocessing, we can compute each of these values in $O(1)$ time. We compute $wpsum(i)$=the sum of the weights from point $1$ to point $i$ ($wpsum(0)=0$ and $wpsum(i \geq 1)=w(i)+wpsum(i-1)$), $wdpsum(i)$=the sum of the values $w(j) \cdot x(j)$ ($1 \leq j \leq i$): $wdpsum(0)=0$ and $wdpsum(i \geq 1)=w(i) \cdot x(i)+wdpsum(i-1)$ (obviously, all of these values can be computed in $O(n)$ time overall). Then, $wsum(i,p)=wpsum(p)-wpsum(i-1)$, $wdsumLR(i,p)=wsum(i,p) \cdot x(p) - (wdpsum(p)-wdpsum(i-1))$, and $wdsumRL(i,p) = (wdpsum(p)-wdpsum(i-1))-wsum(i,p) \cdot x(i)$. With these values, we can binary search the largest value $r$ between $i$ and $j$, such that $wsum(r+1, j)-wsum(i,r-1)>0$ (or $r=i$ if the condition is never met). The optimal location $r_{opt}$ of the median is either $r$ or $r+1$ (if $r+1$ is inside the interval $[i,j]$). The cost of placing the median at position $r_{opt}$ is $wdsumLR(i,ropt)+wdsumRL(ropt,j)$.

This method can be extended to multiple dimensions, as follows. Let's assume that we have a data cube with $d$ dimensions, having $m(j)$ distinct coordinates in each dimension $j$ ($1 \leq j \leq d$) (thus, the cube contains $m(1) \cdot m(2) \cdot \ldots \cdot m(d)$ data points). The $p^{th}$ value in the $j^{th}$ dimension has an assigned coordinate, $x(j,p)$ ($1 \leq j \leq d$; $1 \leq p \leq m(j)$). Each value $Cube(c(1), …, c(d))$ is the weight of a point with coordinates $(x(1,c(1)), …, x(d, c(d)))$. We want to answer efficiently queries of the following type: find the location of the median (under the $L_1$ metric) of all the points with the coordinates in the range $[a(1), b(1)]x...x[a(d), b(d)]$ (a point $(c(1), …, c(d))$ is within the range if $x(j,a(j)) \leq x(j,c(j)) \leq x(j,b(j))$, for every $1 \leq j \leq d$).

This multidimensional range weighted median problem can be reduced to $d$ 1D median problems. The $j^{th}$ such problem considers the range $[a(j),b(j)]$ among $m(j)$ distinct points with appropriately chosen weights $w(j,i)$ ($1 \leq i \leq m(j)$). Let's assume that the median of the $j^{th}$ problem is located at $x(j,r(j))$. The median of all the points in the d-dimensional range is located at $(x(1,r(1)), …, x(d,r(d)))$. In order to answer range weighted median queries, we will first construct $d$ data cubes with the same size as the initial $Cube$: $DCube_j(c(1), …, c(d))=x(j,c(j)) \cdot Cube(c(1), …, c(d))$. For each such data cube $j$ ($1 \leq j \leq d$), we will construct a prefix sum data cube: $PSDCube_j(b(1), …, b(d))$=the sum of the values $DCube_j(c(1), …, c(d))$, with $1 \leq c(j) \leq b(j)$ (for every $1 \leq j \leq d$). As was shown in [6], each prefix sum data cube can be computed in $O(n^d \cdot d)$ time.

Let's denote by $RangeSum(X, [u(1), v(1)], …, [u(d), v(d)])$ the sum of the values $X(c(1), …, c(d))$, with $u(j) \leq c(j) \leq v(j)$ ($1 \leq j \leq d$); if some $v(j)=0$, then the sum is $0$. Using $PSDCube_j$, we can compute $RangeSum(Dcube_j, [u(1), v(1)], …, [u(d), v(d)])$ in $O(2^d)=O(1)$ time. We will also compute a prefix sum data cube for the initial cube: $PSCube(b(1), …, b(d))$ is the sum of all the values $Cube(c(1), …, c(d))$, with $1 \leq c(j) \leq b(j)$ (for every $1 \leq j \leq d$).

With these data cubes, we can solve the range weighted median problem for every dimension $j$ ($1 \leq j \leq d$) as follows. For the $j^{th}$ dimension, the interval $[a(j), b(j)]$ is the 1D interval of points on which we will focus; this interval has $b(j)-a(j)+1$ points, each point $p$ ($a(j) \leq p \leq b(j)$) having the coordinate $x(j,p)$ and a weight $w(p)=RangeSum(Cube, [u(k), v(k)] (1 \leq k \leq d))$, where $u(j)=v(j)=p$ and $u(j')=a(j')$ and $v(j')=b(j')$ (for $1 \leq j' \leq d$, $j' \neq j$). We now have to show how to compute the values $wsum(i,p)$, $wdsumLR(i,p)$ and $wdsumRL(i,p)$. $wsum(i,p)$ is $RangeSum(Cube, [u(k), v(k)] (1 \leq k \leq d))$, where $u(j)=i$, $v(j)=p$, and $u(j')=a(j')$ and $v(j')=b(j')$ (for $1 \leq j' \leq d$, $j' \neq j$). Let's denote by $wdsum(i,p)=RangeSum(DCube_j, [u(k), v(k)] (1 \leq k \leq d))$, where $u(j)=i$, $v(j)=p$, and $u(j')=a(j')$ and $v(j')=b(j')$ (for $1 \leq j' \leq d$, $j' \neq j$). We have: $wdsumLR(i,p)= wsum(i,p) \cdot x(j,p)-wdsum(i,p)$ and $wdsumRL(i,p)= wdsum(i,p)-wsum(i,p) \cdot x(j,i)$. We can compute every value $wsum(i,p)$, $wdsumLR(i,p)$ and $wdsumRL(i,p)$ in $O(2^d)=O(1)$ time. Thus, we can find $r(j)$ in $O(log(n))$ time. The overall time complexity is $O(d \cdot log(n))=O(log(n))$ (since $d$ is a constant).

### 3.4 Multidimensional Range Selection

We consider $d$ sorted arrays $w(1), …, w(d)$, of size $n$ ($w(i,j) \leq w(i,j+1)$, $1 \leq i \leq d$, $1 \leq j \leq n-1$). We want to select the $k^{th}$ smallest weight in the set of points with coordinates $(c(1), …, c(d))$ ($1 \leq c(i) \leq n$, $1 \leq i \leq d$), where the weight of a point $(c(1), …, c(d))$ is $(w(1,c(1))$ *op* $w(2, c(2))$ *op* … $w(d,c(d)))$, where *op* is + (addition), * (multiplication) or *max*. If $O(n^d)$ storage is available, we could store all the point weights and then select the $k^{th}$ smallest weight in $O(n^d \cdot (d+log(n^d)))$, by sorting the weights, or in $O(n^d \cdot d)$ time, by using *QuickSelect*. Instead, we will binary search the $k^{th}$ smallest weight $wk$ in the range $[0,WMAX=(w(1,n)$ *op* … *op* $w(d,n))]$. The feasibility test for a candidate weight $wt$ consists of computing the number of points $p$ whose weight is at most $wt$. if $p \geq k$, then $wt \geq wk$; otherwise, $wt<wk$. Computing $p$ is easy when *op=max*. For each $i$ ($1 \leq i \leq d$) we compute $limit(i)$=the largest index $j$ ($0 \leq j \leq n$) such that $w(i,j) \leq wt$ (using binary search). We consider $w(i,0)=-\infty$. Then, $p=limit(1) \cdot … \cdot limit(d)$. The time complexity of the feasibility test is $O(d \cdot log(n))$. The other cases can be solved by the following recursive function:

**ComputeP(di, wt):**
**if** *(di=1)* **then** {
*binary search the largest index j ($0 \leq j \leq n$), s.t. $w(1,j) \leq wt$*
**return** *j* } **else** { // di$\geq$2
*p=0;* **for** *j=1* **to** *n* **do** *p=p+***ComputeP***(di-1, wt **op**$^{-1}$ w(di, j))*
**return** *p* }

*op*$^{-1}$ denotes the inverse operation of *op* (i.e. *op*$^{-1}$=- for *op=+*, and *op*$^{-1}$=/ for *op=\**). The time complexity of the feasibility test for *op=+* or * is $O(n^{d-1} \cdot log(n))$. The overall complexity of the algorithm is obtained by multiplying the complexity of the feasibility test by *log(WMAX)*. When the weights are integers, the algorithm finds the exact solution. In case of real numbers, it finds the $k^{th}$ smallest weight with any fixed arbitrary precision $\varepsilon>0$ (the binary search ends when the length of the search interval is smaller than $\varepsilon$). If $O(n^q)$ storage is available *($1 \leq q \leq d/2$)*, then we could use the following technique. We consider the first *q* dimensions and generate and store the (multi)set $S_1$ of all the $O(n^q)$ "sums": *(w(1,c(1)) op w(2, c(2)) op … op w(q, c(q)))*, with *$1 \leq c(i) \leq n$, $1 \leq i \leq q$*. Then, we sort all these "sums" and denote by $S_1(j)$ the $j^{th}$ smallest sum in $S_1$. Afterwards, we generate sequentially each of the $O(n^{d-q})$ "sums" considering the dimensions *q+1, …, d*. For each such "sum" *S=(w(q+1,c(q+1)) op w(q+2, c(q+2)) op … op w(d,c(d)))*, with *$1 \leq c(i) \leq n$, $q+1 \leq i \leq d$*, we binary search the largest index *j*, such that $S_1(j) \leq (wt$ *op*$^{-1}$ *S)* (*j* may be *0*), where *wt* is the candidate weight. The sum of all these indices *j* is the number *p* used by the feasibility test. The time complexity of this approach is $O(n^{max\{q,d-q\}} \cdot log(n^q))$. *ComputeP* is a particular case of this more general solution, with *q=1*.

Let's denote by *ww(k)* the $k^{th}$ smallest weight among all the points *($1 \leq k \leq n^d$)*. We now want to be able to compute efficiently an aggregate *agg* of the *k* smallest weights, where *agg=op=+, \** or *max*. When *agg=max*, the answer is *ww(k)*, since *ww(i)$\leq$ww(i+1)* *($1 \leq i \leq n^d-1$)*. For the other cases, we could compute all the weights *ww(1), …, ww(k)*, but this would be too inefficient. We will use the same binary search algorithm as before, but we also compute the prefix "aggregate" array $PS_1$, such that $PS_1(j)=(S_1(1)$ *agg* $S_1(2)$ *agg* *…* *agg* $S_1(j))$. We have $PS_1(j)=PS_1(j-1)$ *agg* $S_1(j)$ (for *j$\geq$1*) and $PS_1(0)=0$ for *agg=+* and *1* for *agg=\**. We will compute a value *pagg*, initialized to *0* (for *agg=+*) or *1* (for *agg=\**). Within the feasibility test, when computing for each "sum" *S* of the remaining $O(n^{d-q})$ "sums" the largest index *j* such that *($S_1(j)$ agg S)$\leq$wt*, we increment *p* by *j* and we set *pagg* to *(pagg agg $PS_1(j)$ agg mop(S, j))*, where *mop(u,v)=u$\cdot$v* for *agg=+*, and *mop(u,v)=$u^v$* for *agg=\**. At the end of the binary search, we obtain the weight *ww(k)*, together with the corresponding values *p* and *pagg*. It is possible that *p$\neq$k*, when the weights are not distinct. The aggregate of the *k* smallest weights will be *(pagg agg mop(ww(k), k-p))*.

## 4 Conclusions

In this paper we presented several novel multidimensional data structures and techniques for computing range aggregate queries efficiently in multidimensional databases (modeled as data cubes). The techniques reduce the time complexity dramatically, compared to the naive solutions. The proposed methods were thoroughly analyzed, mostly from a theoretical perspective. They have many applications in a wide range of domains, as was shown in Section 2.

Efficient range query techniques and data structures were developed in the context of OLAP data cubes [6], multidimensional databases, data compression [7], and even job and communication scheduling [5]. Computing the median of a set of points has applications in personnel scheduling problems [9]. As mentioned in the introduction, the presented techniques can be used in order to efficiently provide accurate input data to decision making optimization techniques [1, 2, 3].

*References:*
[1] C. Resteanu, M. Somodi, M. Andreica, E. Mitan, Distributed and Parallel Computing in MADM Domain using the OPTCHOICE Software, *Proc. of the $7^{th}$ WSEAS Intl. Conf. on Applied Computer Science*, 2007, pp. 376-384.
[2] M. Stoica, D. Nicolae, M. A. Ungureanu, A. Andreica, M. E. Andreica, Fuzzy Sets and Their Applications, *Proc. WSEAS Intl. Conf. on Math. and Comp. in Business. and Econ.*, 2008, pp. 197-202.
[3] H. Ben-Ameur, M. Breton, J. M. Martinez, Dynamic Programming Approach for Valuing Options in the GARCH Model, *Management Science*, Vol. 55, No. 2, 2009, pp. 252-266.
[4] M. A. Bender, M. Farach-Colton, The LCA Problem revisited, *Proc. of the Latin American Symp. on Theoretical Informatics*, *Lecture Notes in Computer Science*, vol. 1776, 2000, pp. 88-94.
[5] M. I. Andreica, N. Tapus, Efficient Data Structures for Online QoS-Constrained Data Transfer Scheduling, *Proc. of the $7^{th}$ IEEE Intl. Symp. on Parallel and Distrib. Computing*, 2008, pp. 285-292.
[6] C.-T. Ho, R. Agrawal, N. Megiddo, R. Srikant, Range Queries in OLAP Data Cubes, *ACM SIGMOD Record*, Vol. 26, No. 2, 1997, pp. 73-88.
[7] P. M. Fenwick, A New Data Structure for Cumulative Frequency Tables, *Software – Practice and Experience*, Vol. 24, No. 3, 1994, pp. 327-336.
[8] R. Fleischer, M. J. Golin, Y. Zhang, Online Maintenance of k-Medians and k-Covers on a Line, *Algorithmica*, Vol. 45, 2006, pp. 549-567.
[9] M. I. Andreica, R. Andreica, A. Andreica, Minimum Dissatisfaction Personnel Scheduling, *Proc. of the $32^{nd}$ ARA Congress*, 2008, pp. 459–463.